# A pitfall in the use of extended likelihood for fitting fractions of pure samples in a mixed sample

A.Nappi

University of Perugia and INFN, Sezione di Perugia

**Abstract**

The paper elucidates, with an analytic example, a subtle mistake in the application of the extended likelihood method to the problem of determining the fractions of pure samples in a mixed sample from the shape of the distribution of a random variable. This mistake, which affects two widely used software packages, leads to a misestimate of the errors.

## Introduction

In particle physics experiments it is often necessary to determine the fractions of several types of events contributing to a measured sample, on the basis of the shape of the distribution of a (possibly multi-dimensional) random variable. The formulation of this problem with the maximum likelihood method is discussed in textbooks about statistical data analysis. The case discussed is the one in which no model relates the expectation value of the total number of events with the fractions of each species, as it would be the case, e.g., if they could be written as functions involving a common set of parameters. In this case, using the total number of events in an additional term, involving its expectation value (extended likelihood), does not add any information, but it may be useful, since it leads to a more symmetric analytic formulation of the problem, with equivalent results[1].

In this paper I want to bring attention to a misunderstanding of the role of the total number of events in the extended likelihood approach, which may lead to a subtle mistake in the error calculations. This is the case, for example, of `HMCMLL`[2], and `TFractionFitter`[3], two widely used software packages that treat the case[4] where the probability density functions (*pdf*) are not specified analytically, but estimated by a Monte Carlo calculation.

The purpose of the paper is to spread awareness on this possible mistake, which may cause incorrect physics results. The approach will be rather didactic and the emphasis will be in trying to elucidate the origin of the mistake in an example where the relevant arguments are not obscured by the complications in the algebra. The case considered is the one of binned data with the *pdf* of the different components specified analytically, but the conclusions are easily extended to the more realistic case, like the one addressed in [4], where information about the *pdf* is provided by Monte Carlo simulations. The well understood issue of the relation between the two, seemingly alternative, approaches of considering the histograms poissonian or multinomial variables, is also discussed, for completeness, in the first section.

Most of the paper discusses the mistake in abstract terms, without referring to explicit implementations. Some information about its effects on the results of the packages quoted earlier, is provided in the last section.

## *1. The maximum likelihood approach*

We shall assume that there is a *mixed* sample under study, containing unknown fractions of two or more *pure* samples. For the latter, the *pdf* of a (possibly multi-dimensional) variable is fully specified.

Let $q_k^{(s)}$ be the binned *pdf* of the pure sample *s* (i.e. the probability that an event of the pure sample *s* falls in bin *k*) and $q_k$ the binned *pdf* of the mixed sample. The model is specified by

$$q_k = \sum_{s=1}^{S} p_s q_k^{(s)} \qquad (1)$$

where $p_s$ are the fractions that one wishes to determine, constrained by

$$\sum_{s=1}^{S} p_s = 1 \quad . \qquad (2)$$

If one assumes that the total number of events in the mixed sample does not carry any information relevant to the problem, it can be treated as a fixed number ( i.e. not a random variable ), in the formulation of the likelihood. Thus, if we call $n_k$ the observed numbers of events in bin *k* in the mixed sample, the distribution of the $n_k$'s is the multinomial distribution

$$P(n_k) = \frac{N!}{\prod_{k=1}^{K} n_k!} \prod_{k=1}^{K} q_k^{n_k} \quad \text{where} \quad N = \sum_{k=1}^{K} n_k \qquad (3)$$

and the problem can be formulated as a maximization of

$$\ln L = \sum_{k=1}^{K} n_k \ln\left(\sum_s p_s q_k^{(s)}\right)$$

with respect to the $p_s$, subject to the constraint $\sum_s p_s = 1$.

The choice to treat *N* as a fixed number, rather than as a random variable, is justified when the total number of events is not related with the fractions $p_s$, e.g. by means of some theoretical model where it can be expressed as a function of common parameters. It can be shown that, under this hypothesis, the conclusion that the total number of events (*N*) does not carry any information on the $p_s$ can be extended to an alternative approach, where *N* is considered a random variable with Poisson distribution. The relation between the two approaches is displayed most clearly by writing the probability distribution of the observations *N*,$n_k$ in terms of the poissonian probability to observe *N* events times the multinomial probability of observing $n_k$ events in the individual bins, conditional to the hypothesis that the total number of events in all bins is *N*

$$P(n_k, N) = P(n_k|N)P(N) = \frac{N!}{\prod_{k=1}^{K} n_k!} \left(\prod_{k=1}^{K} q_k^{n_k}\right) \frac{\nu^N e^{-\nu}}{N!} \qquad (4)$$

containing the additional parameter $\nu$, the expectation value of the total number of events.

This can be used to formulate an *extended likelihood*, function of the parameters $p_s$ ($S-1$ parameters because of the constraint) and of $\nu$:

$$\ln L = \sum_{k=1}^{K} n_k \ln\left(\sum_s p_s q_k^{(s)}\right) + N \ln \nu - \nu \qquad (5)$$

It is useful to stress, at this point, that there are $S$ parameters to be determined: $S-1$ are the fractions that can be fixed independently and one is the expectation value of the total number of events. Assuming that $\nu$ and $p_s$ cannot be expressed as a function of common parameters, they are not mixed in the likelihood function and one obtains the trivial result that the $p_s$ estimates are those that would be determined with the multinomial approach, whereas the maximum likelihood estimate of $\nu$ is given by $\hat{\nu} = N$.

The extended likelihood can be put in a form more useful for applications, using the fact that formula (4) implies that the numbers of events in each bin can be considered independent poissonian variables. By trivial algebra it can be shown, in fact, that

$$P(n_k, N) = P(n_k|N)P(N) = \frac{N!}{\prod_{k=1}^{K} n_k!}\left(\prod_{k=1}^{K} q_k^{n_k}\right)\frac{\nu^N e^{-\nu}}{N!} = \prod_{k=1}^{K} \frac{\left(\sum_s \nu p_s q_k^{(s)}\right)^{n_k} e^{-\sum_s \nu p_s q_k^{(s)}}}{n_k!} = \prod_{k=1}^{K} \frac{(\nu_k)^{n_k} e^{-\nu_k}}{n_k!}$$

with $\nu_k = \sum_s \nu p_s q_k^{(s)}$. $\qquad (6)$

When the problem is formulated in this way, the likelihood function takes the form

$$\ln L = \sum_{k=1}^{K} [n_k \ln \nu_k - \nu_k] \quad \text{with } \nu_k = \sum_s \nu p_s q_k^{(s)}, \qquad (7)$$

which depends on $S$ independent parameters, that can be labeled, e.g.,

$$\nu^{(s)} = \nu p_s \qquad (8)$$

representing the expected numbers of events from sample $s$ in the mixed sample. These parameters are no longer constrained explicitly, since they represent the numbers observed for a given running time and therefore can fluctuate independently.

Apparently no trace is left of the constraint (2). However, the previous discussion shows that the likelihood defined by formulae (7) and (8) can be obtained by a one to one transformation of variables from the likelihood (5), namely

$$\begin{cases} \nu^{(1)} = \nu p_1 \\ \nu^{(2)} = \nu p_2 \\ ... \\ \nu^{(S-1)} = \nu p_{S-1} \\ \nu^{(S)} = \nu\left[1 - \sum_{s=1}^{S-1} p_s\right] \end{cases} \Leftrightarrow \begin{cases} p_1 = \dfrac{\nu^{(1)}}{\sum_{s=1}^{S} \nu^{(s)}} \\ p_2 = \dfrac{\nu^{(2)}}{\sum_{s=1}^{S} \nu^{(s)}} \\ ... \\ p_{S-1} = \dfrac{\nu^{(S-1)}}{\sum_{s=1}^{S} \nu^{(s)}} \\ \nu = \sum_{s=1}^{S} \nu^{(s)} \end{cases}$$

and therefore the values of the $\nu^{(s)}$ that maximize it, can be expressed just replacing in the last formulae the parameters $\nu$, $p_s$ that maximize (5). This implies that the maximization of (7) with respect to the parameters (8), automatically satisfies the normalization condition $\sum_s \hat{v}^{(s)} = N$. This could be proven by direct inspection.

## 2. The errors

If the problem is formulated with the multinomial approach, the covariance matrix of the estimates, $\hat{p}_s$, of the $S-1$ selected $p_s$ can be estimated, as usual, as the inverse of the $(S-1)\times(S-1)$ matrix of second derivatives of the likelihood function

$$\left[-\frac{\partial^2 \ln L}{\partial p_l \partial p_m}\right]^{-1}.$$

Here, as in all subsequent occurrences of second derivatives of the likelihood, it is implied that they are evaluated at the likelihood maximum.

Note the asymmetric treatment of the $p_s$, due to the fact that one of them can be computed as a function of all others using constraint (2). One can, of course, augment the error matrix with an additional row and column, using error propagation on

$$\hat{p}_S = 1 - \sum_{s=1}^{S-1} \hat{p}_s$$

but the error matrix thus obtained is singular.

If the problem is formulated with the poissonian approach, with the likelihood function (7),(8), the parameters estimated are not the fractions but the expected numbers of events[*]. One can estimate the $S\times S$ covariance matrix of the estimates $\hat{v}^{(s)}$ as

$$\left[-\frac{\partial^2 \ln L}{\partial v^{(l)} \partial v^{(m)}}\right]^{-1}.$$

The fractions can be computed a posteriori as

$$\hat{p}_s = \frac{\hat{v}^{(s)}}{\sum_{s=1}^{S} \hat{v}^{(s)}} \qquad (9)$$

and their covariance matrix using error propagation. Note that, if one does that, again the covariance matrix is singular, since definitions (9) are redundant.

## 3. The pitfall

Since the estimates of the $v^{(s)}$ satisfy

$$\sum_{s=1}^{S} \hat{v}^{(s)} = N,$$

one could be tempted to rewrite the likelihood (7) as a function of the $p_s$, setting $v=N$ in (8). One could then perform the minimization directly in terms of the $p_s$ ($S$ parameters!) and obtain the error matrix directly, with no need to perform error propagation on the basis of (9). Doing that, one

---

[*] In the likelihood (7) $v$ and $p_s$ always appear in the combination $vp_s$ and it is not possible to determine them separately

obtains the correct estimates for the $p_s$, but the error matrix is wrong. This will be shown below in an explicit example, but it is already apparent from the fact that this $S \times S$ error matrix is not singular.

Before coming to that example, it will be useful to show that the multinomial and the poissonian approach give consistent results, instead. I will perform an explicit calculation for the case $S=2$, where there is only one non trivial parameter, e.g. $p_1$.

For the multinomial case

$$\ln L = \sum_{k=1}^{K} n_k \ln\left[p_1 q_k^{(1)} + (1-p_1) q_k^{(2)}\right]$$

and the variance of $\hat{p}_1$ is

$$\sigma_M^2(\hat{p}_1) = -\frac{1}{\dfrac{\partial^2 \ln L}{\partial p_1^2}} = \frac{1}{\sum_{k=1}^{K} \dfrac{n_k \left(q_k^{(1)} - q_k^{(2)}\right)^2}{\left[\hat{p}_1 q_k^{(1)} + (1-\hat{p}_1) q_k^{(2)}\right]^2}} \quad (10)$$

For the poissonian approach

$$\ln L = \sum_{k=1}^{K} \left[n_k \ln\left(\nu^{(1)} q_k^{(1)} + \nu^{(2)} q_k^{(2)}\right) - \left(\nu^{(1)} q_k^{(1)} + \nu^{(2)} q_k^{(2)}\right)\right] \quad (11)$$

The covariance matrix of $\hat{p}_1$ and $\hat{p}_2$, using the error matrix of $\hat{\nu}^{(1)}$ and $\hat{\nu}^{(2)}$ and performing error propagation on the basis of formula (9), is given by

$$\begin{bmatrix} \dfrac{\partial \hat{p}_1}{\partial \hat{\nu}^{(1)}} & \dfrac{\partial \hat{p}_1}{\partial \hat{\nu}^{(2)}} \\ \dfrac{\partial \hat{p}_2}{\partial \hat{\nu}^{(1)}} & \dfrac{\partial \hat{p}_2}{\partial \hat{\nu}^{(2)}} \end{bmatrix} \times \begin{bmatrix} -\dfrac{\partial^2 \ln L}{\partial \nu^{(1)2}} & -\dfrac{\partial^2 \ln L}{\partial \nu^{(1)} \partial \nu^{(2)}} \\ -\dfrac{\partial^2 \ln L}{\partial \nu^{(1)} \partial \nu^{(2)}} & -\dfrac{\partial^2 \ln L}{\partial \nu^{(2)2}} \end{bmatrix}^{-1} \times \begin{bmatrix} \dfrac{\partial \hat{p}_1}{\partial \hat{\nu}^{(1)}} & \dfrac{\partial \hat{p}_2}{\partial \hat{\nu}^{(1)}} \\ \dfrac{\partial p_1}{\partial \hat{\nu}^{(2)}} & \dfrac{\partial p_2}{\partial \hat{\nu}^{(2)}} \end{bmatrix} =$$

$$= \frac{1}{\sum_{k=1}^{K}\left[\dfrac{n_k q_k^{(1)2}}{\left(\hat{p}_1 q_k^{(1)} + \hat{p}_2 q_k^{(2)}\right)^2}\right] \sum_{k=1}^{K}\left[\dfrac{n_k q_k^{(2)2}}{\left(\hat{p}_1 q_k^{(1)} + \hat{p}_2 q_k^{(2)}\right)^2}\right] - \left\{\sum_{k=1}^{K}\left[\dfrac{n_k q_k^{(1)} q_k^{(2)}}{\left(\hat{p}_1 q_k^{(1)} + \hat{p}_2 q_k^{(2)}\right)^2}\right]\right\}^2} \times \begin{bmatrix} N & -N \\ -N & N \end{bmatrix} \quad (12)$$

This error matrix shows the expected features: the errors on $\hat{p}_1$ and $\hat{p}_2$ are the same (as they should, since $\hat{p}_2 = 1 - \hat{p}_1$) and their correlation coefficient is 100%. Moreover it can be proven that the diagonal elements of (12) are identically equal to (10). The proof, originally provided by J. Linnemann[5] for an N → ∞ approximation, is given in an appendix for the finite N case. Although the poissonian and multinomial model describe two probabilistically different cases, the identity of the variance estimates is not unexpected, since the extended likelihood (11) can be put in the form of equation (5), which has the same dependence on $\hat{p}_1$ as the likelihood appropriate for the multinomial model.

Conversely, it is easy to prove that the poissonian approach, using the $p$'s as variables, is wrong, as far as errors are concerned. In that case, one has

$$\ln L = \sum_{k=1}^{K}\left[n_k \ln\left(N p_1 q_k^{(1)} + N p_2 q_k^{(2)}\right) - N\left(p_1 q_k^{(1)} + p_2 q_k^{(2)}\right)\right] \quad (13)$$

where $p_1$ and $p_2$ are now parameters to be determined independently, since they are just another name for $\nu^{(1)}$, $\nu^{(2)}$. Their error matrix is given by

$$-\begin{bmatrix} \dfrac{\partial^2 \ln L}{\partial p_1^2} & \dfrac{\partial^2 \ln L}{\partial p_1 \partial p_2} \\ \dfrac{\partial^2 \ln L}{\partial p_1 \partial p_2} & \dfrac{\partial^2 \ln L}{\partial p_2^2} \end{bmatrix}^{-1} = \dfrac{1}{\sum_{k=1}^{K} \dfrac{n_k q_k^{(1)2}}{(\hat{p}_1 q_k^{(1)} + \hat{p}_2 q_k^{(2)})^2} \sum_{k=1}^{K} \dfrac{n_k q_k^{(2)2}}{(\hat{p}_1 q_k^{(1)} + \hat{p}_2 q_k^{(2)})^2} - \left[\sum_{k=1}^{K} \dfrac{n_k q_k^{(1)} q_k^{(2)}}{(\hat{p}_1 q_k^{(1)} + \hat{p}_2 q_k^{(2)})^2}\right]^2} \times$$

$$\times \begin{bmatrix} \sum_{k=1}^{K} \dfrac{n_k q_k^{(2)2}}{(\hat{p}_1 q_k^{(1)} + \hat{p}_2 q_k^{(2)})^2} & -\sum_{k=1}^{K} \dfrac{n_k q_k^{(1)} q_k^{(2)}}{(\hat{p}_1 q_k^{(1)} + \hat{p}_2 q_k^{(2)})^2} \\ -\sum_{k=1}^{K} \dfrac{n_k q_k^{(1)} q_k^{(2)}}{(\hat{p}_1 q_k^{(1)} + \hat{p}_2 q_k^{(2)})^2} & \sum_{k=1}^{K} \dfrac{n_k q_k^{(1)2}}{(\hat{p}_1 q_k^{(1)} + \hat{p}_2 q_k^{(2)})^2} \end{bmatrix}.$$

This matrix does not satisfy the conditions $\sigma^2(\hat{p}_1) = \sigma^2(\hat{p}_2)$, nor the condition of 100% correlation between $\hat{p}_1$ and $\hat{p}_2$. It can also be seen that, at least for some values of $p_1$, the use of these formulae leads to a gross misestimate of the errors. For purpose of comparison with (10), the approximation

$$n_k \cong N(p_1 q_k^{(1)} + p_2 q_k^{(2)}), \ \hat{p}_1 \cong p_1 \ \text{and} \ \hat{p}_2 \cong p_2,$$

valid for N $\to \infty$, is useful. Using the index N to identify results obtained from the likelihood (13), one obtains

$$\sigma_N^2(\hat{p}_1) = \dfrac{1}{N} \dfrac{\sum_{k=1}^{K} \left[\dfrac{q_k^{(2)2}}{p_1 q_k^{(1)} + (1-p_1) q_k^{(2)}}\right]}{\sum_{k=1}^{K} \left[\dfrac{q_k^{(1)2}}{p_1 q_k^{(1)} + (1-p_1) q_k^{(2)}}\right] \sum_{k=1}^{K} \left[\dfrac{q_k^{(2)2}}{p_1 q_k^{(1)} + (1-p_1) q_k^{(2)}}\right] - \left\{\sum_{k=1}^{K} \left[\dfrac{q_k^{(1)} q_k^{(2)}}{p_1 q_k^{(1)} + (1-p_1) q_k^{(2)}}\right]\right\}^2}.$$

The same approximation, in the multinomial case, gives

$$\sigma_M^2(\hat{p}_1) = -\dfrac{1}{\dfrac{\partial^2 \ln L}{\partial p_1^2}} = \dfrac{1}{N} \dfrac{1}{\sum_{k=1}^{K} \dfrac{(q_k^{(1)} - q_k^{(2)})^2}{p_1 q_k^{(1)} + (1-p_1) q_k^{(2)}}}.$$

I computed numeric values for a particular example, where the *pdf*'s of the pure samples are both linear functions of a random variable $x$ contained in $0 \leq x \leq 1$

$$f^{(1)}(x) = 2x$$

$$f^{(2)}(x) = 2(1-x) \ .$$

The results, computed assuming 20 equal size bins, are shown in Fig. 1 as a function of the value of $p_1$.

The fact that likelihood (13) gives the right estimates of the fractions, but a wrong estimate for the errors needs some explanation. If one wants to interpret (13) as a likelihood function, the symbols $p_s$ that appear in it cannot be interpreted directly as the event fractions, but represent just a linear change of variables

$$\nu^{(s)} = N p_s \tag{14}$$

in the expression of the correct likelihood (11).

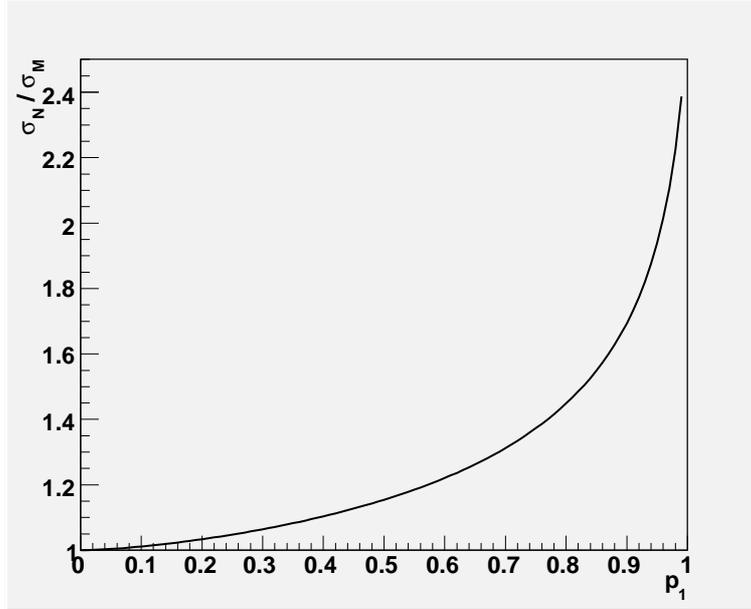

**Figure 1**

**Comparison of the error computed with the incorrect poissonian approach, $\sigma_N(p_1)$, and the multinomial approach, $\sigma_M(p_1)$, in the numerical example considered. The quantity plotted is the ratio of the two errors, computed by the formulae given in the text. By incorrect poissonian approach I mean the one where the expectation values of the numbers of events in the likelihood are written as the measured total number of events times the fractions to be determined**

Such a change of variables is legitimate, whatever the value of the arbitrary constant $N$ is. The values of the $p_s$ that maximize (13), call them $\hat{p}_s$, are related to those of of the $v^{(s)}$ that maximize (11), $\hat{v}^{(s)}$, by

$$\hat{p}_s = \frac{\hat{v}^{(s)}}{N}$$

and cannot be interpreted directly as event fractions. Instead, these will be obtained from formula (9), which, in terms of the $\hat{p}_s$ reads

$$p_s = \frac{\hat{p}_s}{\sum_{s=1}^{S} \hat{p}_s} \,. \tag{15}$$

In the special case where the arbitrary constant N is the measured total number of events, we can use the previous result that, at the likelihood maximum,

$$\sum_{s=1}^{S} \hat{v}^{(s)} = N$$

and, as a consequence, $\sum_{s=1}^{S} \hat{p}_s = 1$ .

However these relations are valid only at the likelihood maximum and therefore it is wrong to assume them for the error calculations that concerns the behaviour of the likelihood away from the maximum, where the $v_s$ are unconstrained parameters.

## 4. Discussion of existing packages.

The `HBOOK`[6] and `root`[3] analysis frameworks, developed at CERN, both include packages (`HMCMLL` and `TFractionFitter` respectively) that implement the method, described in [4], to fit fractions of pure samples in mixed samples, in the case where information about the pure sample *pdf* is provided by Monte Carlo simulations with limited statistics. In this case, Monte Carlo fluctuations are usually accounted for, by treating the integrals over the histogram bins of the *pdf* of the pure samples as additional fit parameters. They are determined, together with event fractions in the mixed data sample, by maximizing a likelihood including terms related to data and Monte Carlo histograms. Reference [4] introduces a clever technique, where maximization with respect to the large number of trivial parameters describing the *pdf*, is performed by a semi-analytical iterative procedure, while maximization with respect to the event fractions is performed by the general minimization package MINUIT[7].

Unfortunately, in reference [4] the likelihood is formulated in terms of an unfortunate choice of the parameters§ related to the event fractions, which corresponds to the $p_s$ of this paper multiplied by the ratio between the statistics of the data sample and the statistics of the Monte Carlo sample *s*. The paper does not clarify the recipe for transforming these parameters into event fractions. However, both in the `HMCMLL` and `TFractionFitter` implementations, the transformation is performed& using the ratio of the total number of events in the data and Monte Carlo samples, rather than their estimates. The two coincide only at likelihood maximum. This mistake is equivalent to the one discussed in this paper, although the algebra is more complicated, due to the fluctuations of the Monte Carlo samples.

A toy Monte Carlo, performed with data distributed according to the *pdf* of the previous numerical example, confirms the results of the analysis presented in the previous sections, when they can be directly compared, i.e. when the contribution of Monte Carlo fluctuations to the error is negligible. In that case, the ratio between the error on $p_1$ computed by `TfractionFitter` and the rms of the estimate of $p_1$, obtained in 1000 Monte Carlo experiments, generated with the same input parameters, is found compatible with $\sigma_N/\sigma_M$ of Fig. 1. Also in cases where the Monte Carlo contribution to the error cannot be neglected, `TfractionFitter` overestimates the error by an amount that depends on the value of $p_1$, with a trend similar to the one of Fig. 1, although with different numerical values. In all cases, a better estimate of the error is obtained propagating the error on formula (15), using the full covariance matrix, retrieved using methods of the class `TfractionFitter`. That it should be so, is obvious when the Monte Carlo contribution to the error is negligible. In that case the constant *N* of formula (14) cancels in the ratio of formula (9), yielding formula (15). No such cancellation occurs, in the general case, for the ratio of the data to Monte Carlo statistics appearing in the definition of the parameters used by `TfractionFitter`.

## 5. Conclusions

Inconsistencies observed in the errors provided by the `HMCMLL` and `TFractionFitter` packages are not related to the formulation of the likelihood, but to an incorrect replacement, in the likelihood function, of expected numbers of events with the total measured number of events times the expected fractions. The extended likelihood approach, based on the use of the Poisson distribution will give the right result if the event fractions are computed by formulae (9).

---

§ The definition is given in the text of ref. [4], between formula (2) and formula (3) on page 219. In their notation, the event fractions, which I call $p_s$, are indicated by $P_s$.

& The transformation used is not mentioned in the documentation, but can be deduced from the code of the method `ComputeFCN`, for `TfractionFitter`, and of the function `HMCLNL`, for `HMCMLL`.

The results provided by these packages are valid for what concerns the estimates of the event fractions, but are incorrect for what concerns the errors, because they are based on the assumption that the normalization condition for the parameters incorrectly interpreted as event fractions, which holds only at the likelihood maximum, is valid everywhere. As a practical remark, the correct errors can be computed from the covariance matrix provided by these packages, applying error propagation to the formula

$$p_s = \frac{\hat{p}_s}{\sum_s \hat{p}_s}$$

Note, however, that the full covariance matrix of the $\hat{p}_s$ must be used in this.

## 6. *Acknowledgements*

I wish to thank S. Goy Lopez and M. Raggi for bringing this problem to my attention and for useful discussions. I am grateful to J.Linnemann for providing stimulating inputs and, in particular, a proof of the identity of the variance estimates in the poissonian and multinomial models.

## *Appendix    Identity of the variance estimates in the multinomial and poissonian models*

Since $\hat{p}_1$ maximizes the multinomial likelihood,

$$\sum_k n_k \frac{q_k^{(1)} - q_k^{(2)}}{\hat{p}_1 q_k^{(1)} + (1-\hat{p}_1) q_k^{(2)}} = 0 \ .$$

Letting $f_k = \hat{p}_1 q_k^{(1)} + (1-\hat{p}_1) q_k^{(2)}$

and expressing $q_k^{(2)}$ as a function of $f_k$ and $q_k^{(1)}$, this yields the identity

$$\sum_k n_k \frac{q_k^{(1)}}{f_k} = N \ . \tag{16}$$

Eliminating $q_k^{(2)}$, the terms of the sum in formula (10) can be written

$$n_k \frac{(q_k^{(1)} - q_k^{(2)})^2}{f_k^2} = n_k \frac{(q_k^{(1)} - f_k)^2}{(1-\hat{p}_1)^2 f_k^2} = \frac{1}{(1-\hat{p}_1)^2}\left[n_k \frac{q_k^{(1)2}}{f_k^2} + n_k - 2 n_k \frac{q_k^{(1)}}{f_k}\right].$$

Letting $A = \sum_k n_k \frac{q_k^{(1)2}}{f_k^2}$ and using identity (16) one can write the inverse of formula (10)

$$I_M = \frac{1}{\sigma_M^2(\hat{p}_1)} = \frac{A - N}{(1-\hat{p}_1)^2} \ .$$

Going now to formula (12), one can proceed similarly, to evaluate the different sums that appear in it. The first one is given by

$$S_1 = \sum_k \frac{n_k q_k^{(1)2}}{f_k^2} = A \ .$$

The second one can be rewritten, eliminating $q_k^{(2)}$ and using again identity (16)

$$S_2 = \sum_k \frac{n_k q_k^{(2)2}}{f_k^2} = \frac{1}{(1-\hat{p}_1)^2} \sum_k n_k \left(1 + \hat{p}_1^2 \frac{q_k^{(1)2}}{f_k^2} - 2\hat{p}_1 \frac{q_k^{(1)}}{f_k}\right) = \frac{N + \hat{p}_1^2 A - 2\hat{p}_1 N}{(1-\hat{p}_1)^2}.$$

Similarly, for the third sum

$$S_3 = \sum_k \frac{n_k q_k^{(1)} q_k^{(2)}}{f_k^2} = \frac{1}{1-\hat{p}_1} \sum_k n_k \left(\frac{q_k^{(1)}}{f_k} - \hat{p}_1 \frac{q_k^{(1)2}}{f_k^2}\right) = \frac{N - \hat{p}_1 A}{1-\hat{p}_1}.$$

Putting all together,

$$I_P = \frac{1}{\sigma_P^2(\hat{p}_1)} = \frac{S_1 S_2 - S_3^2}{N} = \frac{A - N}{(1-\hat{p}_1)^2} = I_M,$$

which proves the identity.